\documentclass[sigchi]{acmart}
\usepackage{caption}
\usepackage{float}
\renewcommand\footnotetextcopyrightpermission[1]{} 

\acmConference[CHI Play 2020]{The international and interdisciplinary conference for researchers and professionals across all areas of play, games and human-computer interaction.}
{November 2-4, Ottawa, Canada}

\begin{document}

\title{crea.blender: A Neural Network-Based Image Generation Game to Assess Creativity}

\author{Janet Rafner}
\email{janetrafner@phys.au.dk}
\affiliation{
  \institution{Aarhus University}
  \city{Aarhus}
  \country{Denmark}
}

\author{Arthur Hjorth}
\email{arthur@phys.au.dk}
\affiliation{%
  \institution{Aarhus University}
  \city{Aarhus}
  \country{Denmark}
}

\author{Sebastian Risi}
\email {sebr@itu.dk}
\affiliation{%
 \institution{IT University of Copenhagen}
 \city{Copenhagen}
 \country{Denmark}
}

\author{Lotte Philipsen}
\email{lottephilipsen@cc.au.dk}
\affiliation{%
  \institution{Aarhus University}
  \city{Aarhus}
  \country{Denmark}
}

\author{Charles Dumas}
\email{charles97.dumas@gmail.com}
\affiliation{%
  \institution{Aarhus University}
  \city{Aarhus} 
  \country{Denmark}
}

\author{Michael Mose Biskjær}
\email{mmb@cc.au.dk}
\affiliation{%
   \institution{Aarhus University}
    \city{Aarhus}
    \country{Denmark}
}

\author{Lior Noy}
\email{lior.noy.1@gmail.com}
\affiliation{%
    \institution{IDC Herzliya and Weizmann Institute}
    \city{Rehovot}
    \country{Israel}
}

\author{Kristian Tyl\'en}
\email{kristian@cc.au.dk}
\affiliation{%
  \institution{Aarhus University}
  \city{Aarhus}
  \country{Denmark}
}

\author{Carsten Bergenholtz}
\email{cabe@mgmt.au.dk}
\affiliation{%
 \institution{Aarhus University}
 \city{Aarhus}
 \country{Denmark}
}

\author{Jesse Lynch}
\email{mrdrjbird@gmail.com}
\affiliation{%
 \institution{Aarhus University}
 \city{Aarhus}
 \country{Denmark}
}

\author{Blanka Zana}
\email{blanka.zana@scienceathome.org}
\affiliation{%
  \institution{Aarhus University}
 \city{Aarhus}
 \country{Denmark}
}

\author{Jacob Sherson}
\email{sherson@phys.au.dk}
\affiliation{%
  \institution{Aarhus University}
  \city{Aarhus}
  \country{Denmark}
 }

\renewcommand{\shortauthors}{Rafner and Sherson, et al.}
\begin{abstract}
We present a pilot study on \textit{crea.blender}, a novel co-creative game designed for large-scale, systematic assessment of distinct constructs of human creativity. Co-creative systems are systems in which humans and computers (often with Machine Learning) collaborate on a creative task. This human-computer collaboration raises questions about the relevance and level of human creativity and involvement in the process. We expand on, and explore aspects of these questions in this pilot study. We observe participants play through three different play modes in \textit{crea.blender}, each aligned with established creativity assessment methods. In these modes, players ``blend'' existing images into new images under varying constraints. Our study indicates that crea.blender provides a playful experience, affords players a sense of control over the interface, and elicits different types of player behavior, supporting further study of the tool for use in a scalable, playful, creativity assessment.    

\end{abstract}



\maketitle

\section{Introduction and Related Work}

Creativity is commonly understood as the combination of novelty and value \cite{runco2012standard}, and is one of the most prized skills of the 21st century \cite{pwc_2017}. Creative processes are explored extensively in the burgeoning field of creativity support tools and co-creative systems \cite{lin2020your, oh2018lead, sethapakdi2019painting, frich2019mapping}. These fields are faced with a fundamental trade-off between imposed constraints, granularity of the problem representation, and user control \cite{csikszentmihalyi2014creative, hewett2005informing}: On one hand, low degrees of automated support leave the user in more control, but typically at the expense of requiring extensive training and/or labor in performing fine-grained operations in the creation of creative products. On the other hand, high levels of automated support may enable rapid production of creative products, but the loss of detailed user control leaves the relevance and level of human creativity and involvement in the process unclear. 

In this paper, we present a new co-creative system, crea.blender and use it to investigate if a ML-based image generation game can provide appropriate, coarse-grained support to allow for playful and scalable assessment of human creativity.

\subsection{Creativity Assessment}

Established methods for measuring creativity often focus on two processes: divergent and convergent thinking. Divergent thinking (DT) is commonly referred to as the process of thinking flexibly and using existing knowledge to come up with new ideas and solutions \cite{kaufman2010cambridge, guilford1968intelligence}. Convergent thinking (CT) is the process of selecting which of those ideas is worth further elaboration \cite{kaufman2010cambridge, guilford1968intelligence}.

The use of games for creativity assessment is picking up traction \cite{shillo2019detecting, hart2017creative, huang2010idea} as it has been shown that game-based psychometric tests can combat test anxiety or the researcher effect, thus providing cleaner data on the tested phenomenon \cite{dicerbo2014game}. Additionally, unlike common DT and CT tests that record only the discovered \textit{solution}, games can record the \textit{process} of exploration and convergence to a solution \cite{hart2017creative}.

crea.blender is intended to be the centerpiece of the online game-based large-scale portfolio, \href{https://hybridintelligence.eu/crea}{CREA} \cite{CREA}, which has been designed in response to the call for portfolio based assessment of creativity \cite{cortes2019re, reiter2019scoring, acar2019divergent}. 

\subsection{ML-Supported Image Generation}
Generative Adversarial Networks (GANs) \cite{goodfellow2014generative} are the most widely-used type of ML models for image generation \cite{bailey2020tools, schneider2018has}. GANs traditionally consist of two competing models: A generator, that is trained to generate images, and a discriminator, that is trained to distinguish between real images, and images created by the generator.
Artists have recently started using just the trained generator to produce images. When used as a tool for artistic expression, one can feed an input vector into the generator and it will provide an image based on both the vector, and its internal state, i.e. its weights and biases. Importantly, while the internal state of the generator directly corresponds with the features of the generated image, these features do not necessarily align with the features that a human would perceive. 

For example, in a picture containing an orange ball and trees in the background, humans intending to enhance a distinct shape from an image (e.g. ‘the orange ball’) might have the frustrating experience that the system instead enhances some blurred trees in the background that the participant didn’t even notice. The question of user control over GANs for image generation is thus fundamental in determining the systems feasibility for creative processes \cite{mazzone2019art, hertzmann2019visual}.

\section{Exploratory Study}
While both manually blended images \cite{shklovsky2016viktor} and computer generated images \cite{crawford2019excavating} have been explored extensively, here we transform a co-creative image generation system into a playful game for the general public.Our game also carefully aligns with established task and game-based creativity assessments \cite{kwon1998developing, beketayev2016scoring, lau2010creativity, shillo2019detecting, hart2017creative, huang2010idea}. However before we can assess DT and CT in \textit{crea.blender}, we need to address the fundamental question: \textit{does crea.blender’s interface support a playful, controllable, and versatile image manipulation user interaction?} Concretely we address 

\begin{enumerate}
    \item {\textbf{ Player control:} to what extent does the interface afford players to intentionally express creativity?  }
    \item {    \textbf{Varying types of behavior:} Do we see participants playing differently in different play modes? }
    \item {\textbf{ Playfulness:} Does this interaction with \textit{crea.blender} make users feel playful? }
\end{enumerate}


\subsection{Presenting crea.blender}

crea.blender affords creativity by letting players ``blend'' existing images into new images. Using BigGAN \cite{brock2018large}, which has been trained on ImageNET \cite{deng2009imagenet}, and by providing sets of between 3-6 source images, players can easily create a large number of new images by simply adjusting how much of each source image will be blended in. \textit{crea.blender} takes inspiration from the project Artbreeder (originally Ganbreeder), which aims to be a new type of creative tool that empowers users creativity and collaboration \cite{simon2019}. \textit{crea.blender} works with one of the core aspects of creativity, constraint-based combinational creativity, which is here conceived visually as a means to achieve a creative outcome \cite{costello2000efficient}. 

\begin{figure}[h]
  \centering
  \includegraphics[scale=0.25]{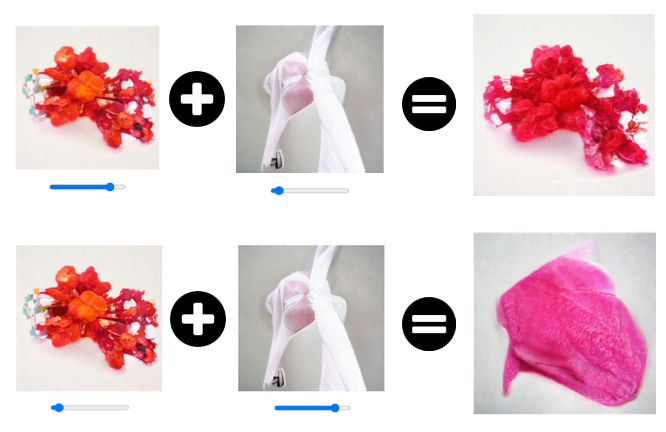}
  \caption[Illustration of \textit{crea.blender's} mechanics]%
  {Illustration of \textit{crea.blender's} mechanics  \normalfont \textit{When mixing images in crea.blender, players use sliders to indicate how much they want each image to contribute to the generated image.  A vector is calculated from these weights and the underlying vector of each respective source image. This new vector is then passed into the GAN. Above we illustrate how two source images can produce two relatively differently looking images, depending on which is weighted higher. }}
  \label{Figure 1}
\end{figure}

crea.blender has three modes, each designed to afford (and test) specific aspects of creativity. Due to the focus of this paper, most discussions of image selection, wording of instructions, timing, etc., are outside the scope of this paper and will only be briefly described.

\begin{figure}
\centering
  \includegraphics[width=\linewidth]{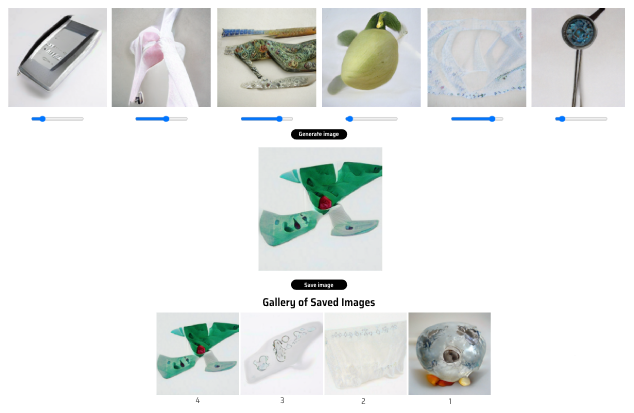}
    \caption[Creatures and Open Play mode interface]%
    {Creatures and Open Play mode interface. \normalfont \textit{ The top row of six images are the source images that players can blend together. Below this, we see the last generated image. At the bottom, we see images that the player has saved.}}
    \label{Figure 2}
\end{figure}

\begin{figure}
\centering
  \includegraphics[width=\linewidth]{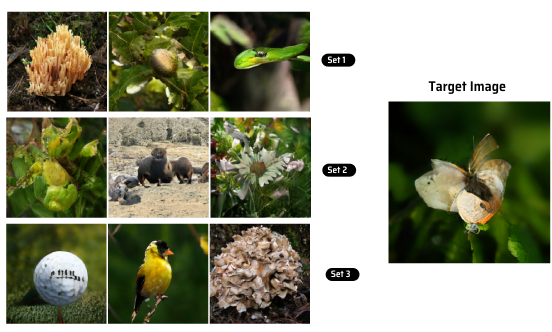}
    \caption[Challenge mode interface.]%
    {Challenge mode interface.  \normalfont \textit{The source image sets are on the left and the  target image is on the right. When a set is chosen, the player blends them similar to the other modes (Figure \ref{Figure 2})}}
    \label{Figure 3}
\end{figure}

\begin{enumerate}
    \item{ \textbf{Creatures:} (Figure \ref{Figure 2}) Players are presented with six images and are asked to create and save as many different ``animal-like'' creatures as possible in five minutes.}
    \item {\textbf{Challenge:} (Figure \ref{Figure 3}) Players are presented with a target image and three sets of three source images. Only one set can produce the exact target image, and players’ objective is first to determine which set was used to create the target image (up to 30 seconds), and then to recreate their closest approximation of the target image (up to three min). There are three levels in the Challenge mode.}
    \item { \textbf{Open Play:} (Figure \ref{Figure 2}) Players are presented with the same six source images as they used in Creatures mode. Unlike in the Creatures mode, they are asked to create \textit{any} image (not just animal-like) they find interesting during five minutes of playtime.}
\end{enumerate}
 
\subsection{Procedure and Data collection}

For the pilot study we convenience sampled and recruited eight participants from our institution.  Participants were asked to Think Aloud \cite{ericcson1978think} while playing with \textit{crea.blender}. Each user session, (including the follow-up user experience survey) took about 40 minutes. While the final version of \textit{crea.blender} will be built in Unity for cross-platform access, we built this prototype in Python3 using the Flask framework, and our participants played with it on a desktop with a mouse at our lab.

Participants were audio recorded, and two researchers were present and wrote observational field notes. crea.blender saved the image, the slider values and a timestamp for each time a player generated an image. 

\subsection{Results and Discussion}
In the following section we address the three themes of the paper by looking presenting parts of the data collected.

\subsubsection{User Control} 
\begin{figure}[h]
  \centering
  \includegraphics[width=\linewidth]{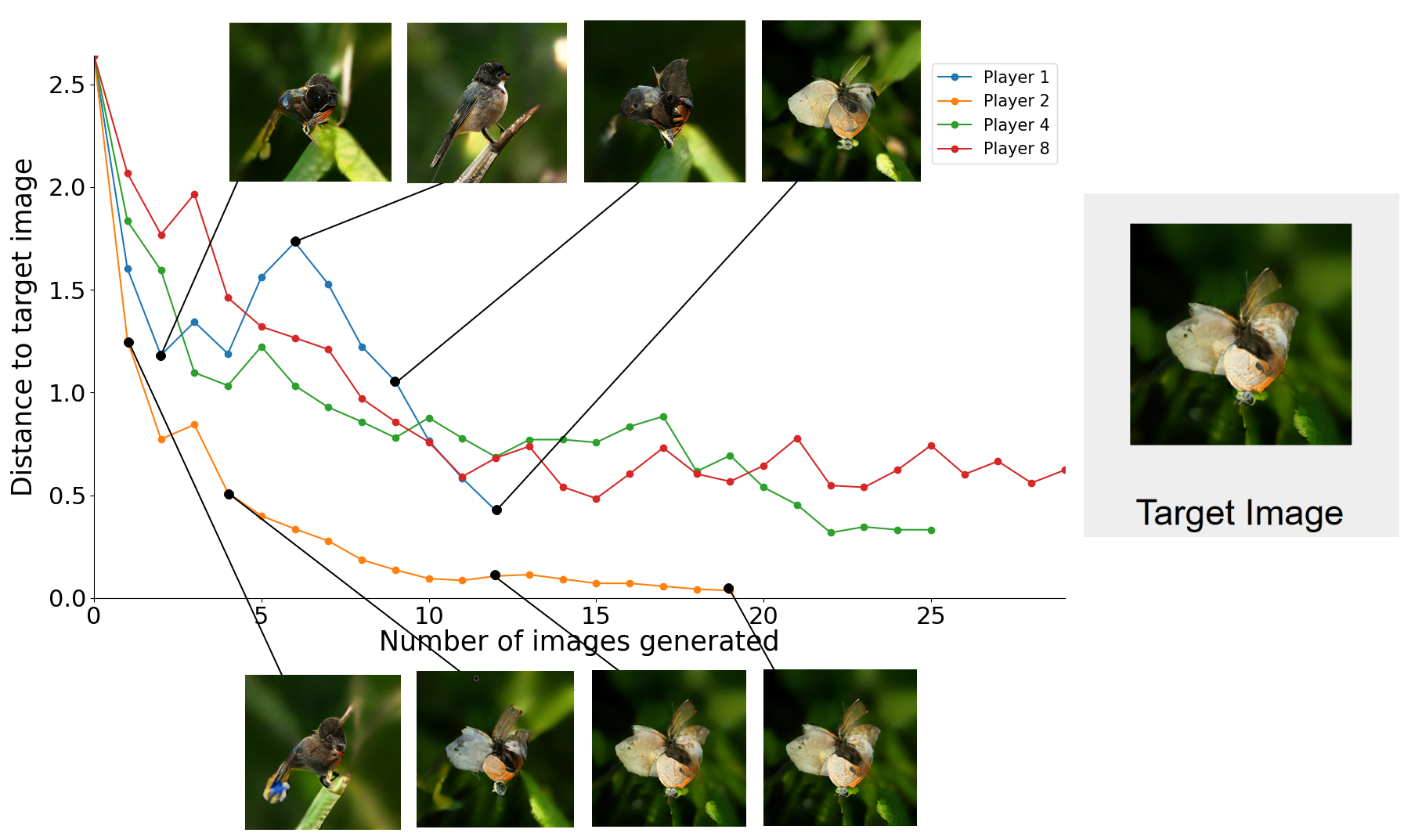}
  \caption[Players’ convergence on target image]%
{Players converge on the target image in Challenge Mode. \normalfont \textit{We see how players consistently get closer to the target image.}}
   \label{Figure 4}
\end{figure}

 We address in three ways players’ feelings of, and ability to exhibit control over crea.blender.  We first present data from the Challenge mode in which players have to generate their closest approximation of a predetermined image. The goal-oriented nature of the task allows us to measure whether players’ interactions with crea.blender were seemingly random or seemingly directed towards the pre-specified goal. Specifically, we can see whether players get closer to the target image in a controlled incremental way or whether they happen to stumble upon it. As a proxy for distance to the target image, in Figure \ref{Figure 4} we plot for each image how far each slider is from the correct setting.

Reading Figure \ref{Figure 4} from left to right shows player progression towards the target. If a player reaches 0 on the y-axis, they have perfectly re-created the target. The orange line (Player 2) shows near-monotonic convergence towards the target-image while the blue line (Player 1) shows more explorative progression. For all players, we see at most two ‘worsenings’ away from the target image before the player corrects their action and moves closer to the target image. This suggests that players recognize when they have gone off track and immediately know how to correct it, and thus are indeed in some sense in control during the Challenge mode.

The second way in which we address user control is by presenting data from the post-play survey. Here, we asked players to indicate on a 1-6 Likert scale how much control they felt they had at the beginning of and at the end of playing, respectively. We conducted a pairwise t-test and calculated the effect size. Based on the results, we can say that there is a large increase (2 to 3.5 on average) in how much control the players felt towards the end of the task, compared to the beginning (Cohen's d = 1.3, p = 0.003).

Finally, we illustrate player control and intentions through a brief vignette from one of our player's Think Aloud transcripts, logs and observations. In this particular vignette, we analyze Player 4 (P4) during Challenge mode after they had just chosen the correct set of source images. P4 immediately turns up all three sliders, one of them by 0.24 and the other two by about 0.4, and generates an image that looks like a bird. P4 then says,

\begin{quote}
Okay. I can only see the color of the beak matching this thing. So let's try. Let's try with that. That may be totally off anyways but, I kind of have this...  this is not quite a ball, but it's maybe close and maybe if you mix a bird with a golf ball you get something close.
\end{quote}

What we see in this quote is that P4 attends to, first, the target image, and looks for colors in the source images that match its colors. P4 then attends to the shape of one of the source images, a golf ball, and hypothesizes that if they mix some more of that shape with their current image, they will get closer to the target image. Around one minute later, P4 produces their tenth image (Figure 6, image 10) which is quite to the target image. P4 now focuses on features in their generated image to fine-tune their creation.

\begin{figure}
  \includegraphics[width = \linewidth]{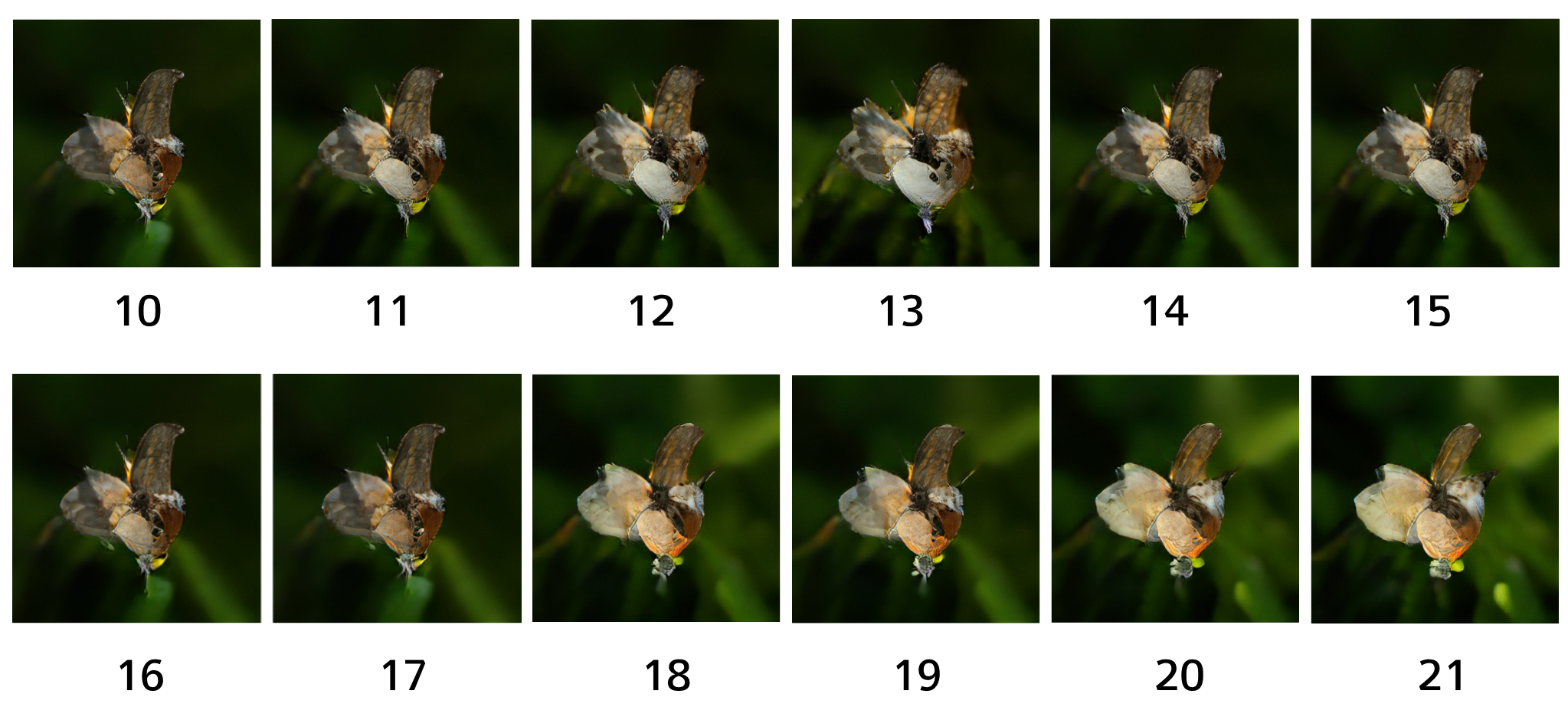}
     \caption[subset of images produced by P4 during challenge mode .]%
    {Subset of images produced by P4 during Challenge mode.  \normalfont \textit{These 12 images, pared with quotes from the P4 to examine if players exhibit control over crea.blender}}
\label{5}
\end{figure}

\begin{quote}
    (\#15). Yeah the left wing now is kind of... “perfect” is a bit too strong word, but it's pretty good. And you also have this slight, slight antenna here (\#16), which is kind of bit to the side (\#17). No (\#18), I don't quite recall how it was made, maybe this one I didn't touch so much. Oh, that's getting very close (\#19). I think because this is growing out... let's go back (\#20). So it looks like when this goes up a bit, this part grows out... and this part gets a bit slimmer (\#21) on the right. 
\end{quote}

Importantly here, P4 does not - at least explicitly - reason about transferring features \textit{from} the source images, but has nonetheless acquired a sense of how different proportions of each source image affects the generated image, and they use this to successfully navigate towards a close approximation of the target image.

These two quotes illustrate a recurring theme across all players: sometimes, players would attend to features (colors, shapes, textures, etc.) in the source images, and hypothesize how mixing them together could produce the target image. Other times, players focused purely on how different slider settings affect features in the generated image. These two gestalts offered complementary perspectives, and together, they enabled participants to generate images relevant to each game mode.

\subsubsection{Varying types of behavior}

\begin{figure}[h]
  \centering
  \includegraphics[width=  \linewidth]{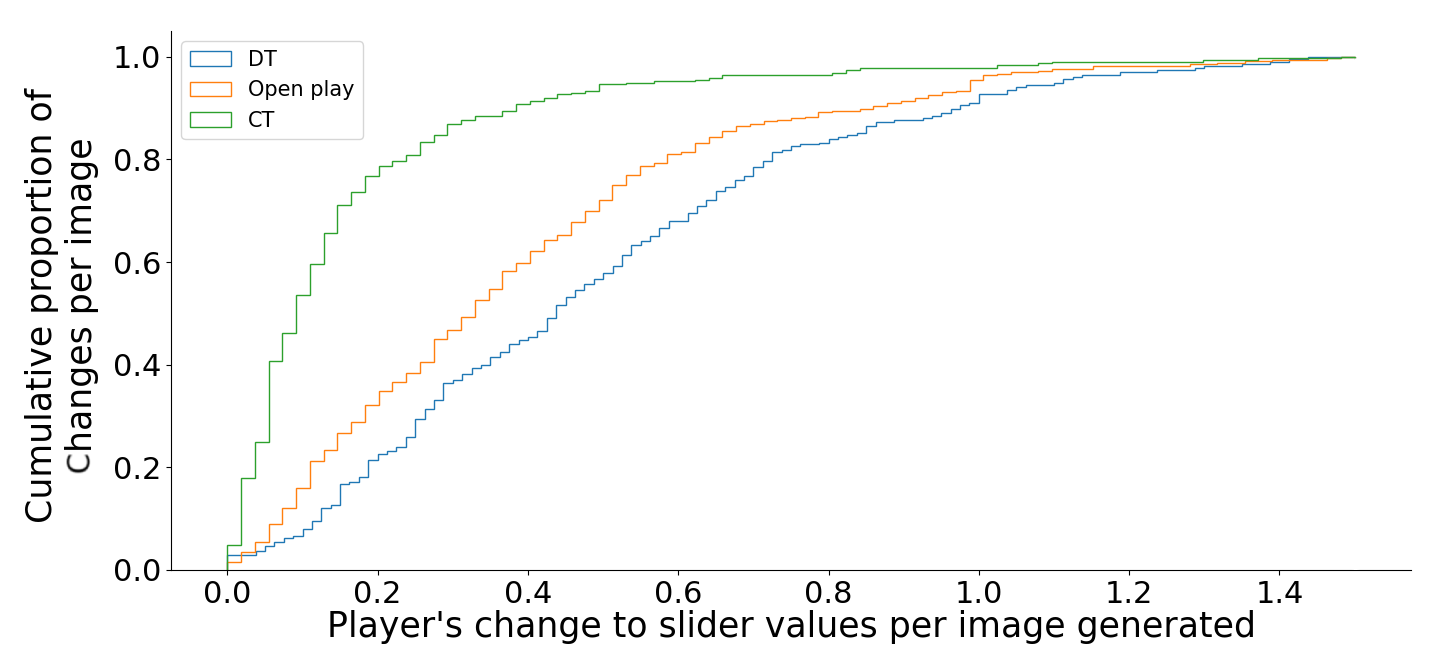}
  \caption[cumulative histogram of slider changes for all players in the different modes]%
      {Cumulative histogram of slider changes for all players in the different modes. \normalfont \textit{}}
   \label{Figure 6}
\end{figure}

It is important to explore if different game-mode prompts in crea.blender can elicit different types of behavior as each mode is tied to specific creative processes. The primary interaction method in crea.blender is changing the weights of each image with the corresponding slider before generating a new image. Therefore, one approach is to look for systematic differences in the size of the changes to sliders players make in the different modes (Figure \ref{Figure 6}).  DT is most commonly associated with an open, explorative process whereas CT is commonly associated with iterative narrowing in on a particular solution or idea \cite{guilford1968intelligence}.  Thus, in the Creatures mode (DT task), we expect much larger average step sizes when creating images than in the challenge mode (CT task)

On average players generated 33.5 (SD$=13$), 86.25 (SD$=29$), and 43.38 (SD$=11$) images in the Creates, Challenge and Open Play modes respectively. Players' changes to sliders ranged from small (iterative) to large (explorative). Figure \ref{Figure 6} shows the cumulative fraction of total changes in sliders for players per image generated. For instance, 78\% of images in the Challenge mode were generated with a change smaller than 0.2. In contrast, 79\% and 66\% of images respectively in the Creatures and Open Play modes were made with changes larger than 0.2. A Kruskal-Wallis test (p<0.001) and post-hoc pairwise Mann-Whitney U tests revealed small but significant differences in behavior between the open play and creatures mode ($\rho$ = 0.17, p <0.001) and much more dramatic differences between these and the challenge mode ($\rho$ =0.66 and  $\rho$ = 0.55, respectively, p<0.001). The latter confirms expectations from previous creativity research, whereas the former provides intriguing input to further work. This demonstration that crea.blender can indeed drive different types of behavior thus fulfills an important criterion for assessing its suitability as a means for assessing creativity.

\subsubsection{Playfulness}
As a final investigation of the suitability of crea.blender as a basis for future creativity research we now turn to the question of perceived playfulness. This is essential for realizing large-scale adoption of the game portfolio. Players were asked on a 1-6-Likert scale to rate how playful they felt overall throughout the game. The mean of their rating was 4.375, with a mode of 4. No one rated below 3, and two out of the eight players rated it as a 6. These data suggest that crea.blender feels like a playful experience and our observations of gameplay have provided indications on how this can be improved.

\section{Conclusion and Outlook}

From this pilot study we can conclude that ML-assisted image generation provides a promising playground for research in creativity assessment. In our data analysis, we found that players were able to intentionally create images with crea.blender; that the game encouraged different uses depending on the creative constraints on the mode; and that players by and large found it playful.  
This pilot was the first step in order to determine if crea.blender is feasible to systematically study creativity in a playful way. Our pilot data support this use, however larger studies must be done to substantiate the work.

We plan to further investigate CT and DT in crea.blender and incorporate crea.blender into the full CREA suite to provide a holistic and scalable approach to testing creativity in a playful way.

\section{Acknowledgments} 
The authors thank the Novo Nordisk, the Synakos, and Carlsberg Foundations for their generous support, and Mads Kock Pedersen for useful discussions.


\end{document}